\documentstyle[11pt,a4]{article}
\newlength{\bredde}
\def\slash#1{\settowidth{\bredde}{$#1$}\ifmmode\,\raisebox{.15ex}{/}
\hspace*{-\bredde} #1\else$\,\raisebox{.15ex}{/}\hspace*{-\bredde} #1$\fi}
\newcommand{\beq}{\begin{equation}}
\newcommand{\eeq}{\end{equation}}
\newcommand{\bea}{\begin{eqnarray}}
\newcommand{\eea}{\end{eqnarray}}
\newcommand{\noi}{\vspace{12pt}\noindent}
\newcommand{\lG}{\raise.3ex\hbox{$\stackrel{\leftarrow}{G}$}}
\newcommand{\lU}{\raise.3ex\hbox{$\stackrel{\leftarrow}{U}$}}
\newcommand{\lP}{\raise.3ex\hbox{$\stackrel{\leftarrow}{{\cal P}}$}}
\newcommand{\leta}{\raise.3ex\hbox{$\stackrel{\leftarrow}{\eta}$}}
\newcommand{\lOmega}{\raise.3ex\hbox{$\stackrel{\leftarrow}{\Omega}$}}
\newcommand{\ldr}{\raise.3ex\hbox{$\stackrel{\leftarrow}{\delta^r}$}}
\newcommand{\ldl}{\raise.3ex\hbox{$\stackrel{\leftarrow}{\delta^l}$}}
\newcommand{\rdr}{\raise.3ex\hbox{$\stackrel{\rightarrow}{\delta^r}$}}
\newcommand{\rdl}{\raise.3ex\hbox{$\stackrel{\rightarrow}{\delta^l}$}}

\def\beqn{\begin{eqnarray}}
\def\eeqn{\end{eqnarray}}

\def\gtwid{\raise.3ex\hbox{$>$\kern-.75em\lower1ex\hbox{$\sim$}}}
\def\ltwid{\raise.3ex\hbox{$<$\kern-.75em\lower1ex\hbox{$\sim$}}}

\begin{document}
\begin{titlepage}
\title{\Large{ BV GAUGE THEORIES}}


\author{{\sc Jorge Alfaro}\\Fac. de Fisica\\Universidad Cat\'{o}lica 
de Chile\\
Casilla 306, Santiago 22, Chile\\
	jalfaro@lascar.puc.cl}

\maketitle
\vfill
\begin{abstract}Using the Non-Abelian Batalin-Vilkovisky formalism
introduced recently, we present a generalization of  the Yang-Mills
gauge transformations , to
include antisymmetric tensor fields as gauge bosons. The Freedman-Townsend
transformation for the two-form gauge field is automatically recovered. 
New characteristic
classes involving this two-form field  and the Yang-Mills one-form field are 
derived. 
We also show how to include, in an unified way,
a gauge invariant coupling of the new gauge bosons to fermionic and bosonic
matter.
\end{abstract}
\vfill
\begin{flushleft}
PUC-2/97\\
\end{flushleft}
\end{titlepage}
\newpage


\section{Introduction}
\noi
Yang-Mills gauge theories have provided the building principle for all the 
known interactions
of nature. They enable the unified description of the weak and 
electromagnetic interactions
as well as the strong interactions, garantying the renormalizability and 
unitarity  of the theory.

Gravity is a gauge theory of a different kind. This difference is precisely
the main reason of why we do not have a theory of Quantum Gravity: In Yang-Mills
gauge theories, the gauge symmetry at our disposal enable the formulation of
unitary and renormalizable models. The gauge symmetry of General Relativity, on
the other hand, is not enough to have renormalizability.

The most promising road to quantize gravity is provided by String 
Theory\cite{witten1}. This
is so because in string theories the amount of gauge symmetries increases 
enormously.
So much that the model is not only renormalizable, but finite.

It is hard to get physical information from string theories. Partially this is due
to the way the theory has been studied up to now. The First Quantization
do not exhibit the symmetries of the interacting theory explicitly and so
part of the beauty is hidden. In fact the Second Quantized version of the model
do exhibit an enormous symmetry \cite{witten2}. However the known formulations
are still background dependent and explicitly refer to the first quantized
version\cite{zwiebach}. Most of the powerful methods of physical analysis 
available in conventional
Quantum Field Theory (instantons, background field method) require a background 
independent
formulation of the theory which exhibit 
explicitly the whole symmetry involved.

For this reason it would be desirable to have a principle to built String Field 
Theories
directly at the Second Quantize level. Such a principle will certainly be 
a much enlarged gauge symmetry principle encompassing Yang-Mills symmetries,
invariance under general coordinate transformations and many other yet unknown 
symmetries.

One of the motivations of this work is to explore new kinds of gauge symmetries
based in the algebraic structure built in the Batalin-Vilkovisky(BV) quantization
method\cite{BV}.

In a series of  papers we have been uncovering a rich algebraic 
structure
that generalizes the Batalin-Vilkovisky method of quantization\cite{AD,2,AD2,AD1,last}.
New nilpotent $\Delta$ operators and generalized antibrackets emerge. 
This induces,
in a standard manner, a master equation and a local symmetry transformation.
The structure
is reminiscent of the one discussed by Sen \cite{sen} and 
Zwiebach\cite{zwiebach} in the context of String Field Theory.

As we mentioned above, it is interesting to consider extensions of the Yang-Mills 
gauge principle
because these new  local symmetries may be a part of the Gauge Principle of String
Field THeory. From the physical point of view, the new gauge fields involved 
in these
generalizations may provide corrections to the Standard Model predictions 
which reflect, at low energies,
the existence of a string at high energies.

Presently, we want to exhibit an explicit realization of this algebraic 
scheme to provide unification of interacting higher forms gauge fields with 
values
on a Lie algebra. 

As a concrete example of the formalism, we construct Characteristic Classes
involving a two form non-abelian gauge field. The transformation of this
field automatically coincides with the Freedman-Townsend transformation
\cite{freedman}
\cite{marc}.

Finally, we explain how to include matter coupled in a gauge invariant
way to these higher form gauge fields.

\section{A review of the Non-abelian BV structure}

\noi
The conventional antibracket of the Batalin-Vilkovisky formalism can be
viewed as being based on a 2nd-order odd differential operator $\Delta$
satisfying $\Delta^2 = 0$. In (super) Darboux coordinates it takes the
simple form \cite{BV}
\beq
\Delta = (-1)^{\epsilon_A+1}\frac{\delta^r}{\delta\phi^A}
\frac{\delta^r}{\delta\phi^*_A} ~,\label{Delta}
\eeq
where to each field $\phi^A$ one has a matching ``antifield'' $\phi^*_A$
of Grassmann parity $\epsilon(\phi^*_A) = \epsilon(\phi^A)+1$.
The antifields are conventional antighosts of the Abelian shift symmetry
that for flat functional measures leads to the most general
Schwinger-Dyson equations \cite{AD}.

\noi
Given $\Delta$ as above, one can define an odd (statistics-changing)
antibracket $(F,G)$ from the failure of $\Delta$ to act like a derivation:
\beq
\Delta(FG) = F(\Delta G) + (-1)^{\epsilon_G}(\Delta F)G
+ (-1)^{\epsilon_G}(F,G) ~. \label{abdef}
\eeq
The antibracket so defined automatically satisfies the following relations.
First, it has an exchange symmetry of the kind
\beq
(F,G) = (-1)^{\epsilon_F\epsilon_G+\epsilon_F+\epsilon_G}(G,F)~.
\label{exchange}
\eeq
It also acts like a derivation in the sense of a generalized Leibniz rule:
\begin{eqnarray}
(F,GH) &=& (F,G)H + (-1)^{\epsilon_G(\epsilon_F+1)}G(F,H) \cr
(FG,H) &=& F(G,H) + (-1)^{\epsilon_G(\epsilon_H+1)}(F,H)G~, \label{Leibniz}
\end{eqnarray}
and it satisies a Jacobi identity,
\beq
\sum_{\mbox{\rm cycl.}}(-1)^{(\epsilon_F+1)(\epsilon_H+1)}(F,(G,H))
= 0~.\label{Jacobi}
\eeq
In addition, there is a useful relation between the $\Delta$-operator and
its associated antibracket:
\beq
\Delta (F,G) = (F,\Delta G) - (-1)^{\epsilon_G}(\Delta F,G) ~.\label{dfg}
\eeq

\noi
Recently \cite{AD1}, it was shown that
the antibracket formalism is open to a natural generalization. In a
path-integral formulation, this generalization can be derived by considering
general field transformations $\phi^A \to g^A(\phi',a)$, where $a^i$
represent certain collective fields \cite{AD2}. The idea is to impose
on the Lagrangian path integral the condition that certain Ward identities
are preserved throughout the quantization procedure. If one imposes the
most general set of Ward identities possible -- the Schwinger-Dyson
equations -- through an unbroken Schwinger-Dyson BRST symmetry \cite{AD3},
one can recover the antibracket formalism of Batalin and Vilkovisky by
integrating out certain ghosts $c^A$ (the antifields $\phi^*_A$ being simply
the antighosts corresponding to $c^A$). For flat functional measures
this corresponds to
local shift transformations of the fields $\phi^A$. If the measure is not
flat, or if one wishes to impose a more restricted set of Ward identities
through the BRST symmetry, the $\Delta$-operator and the associated
antibracket will differ from those of the conventional Batalin-Vilkovisky
formalism. In ref. \cite{AD} it was shown how the Batalin-Vilkovisky
$\Delta$-operator (\ref{Delta}) can be viewed as an Abelian operator
corresponding to the Abelian shift transformation
\mbox{$\phi^A \to \phi^A - a^A$}.
The analogous non-Abelian $\Delta$-operator for general transformations
$\phi^A \to g^A(\phi'^A,a)$ was derived in ref. \cite{AD2}:
\beq
\Delta G \equiv (-1)^{\epsilon_i}\left[\frac{\delta^r}{\delta\phi^A}
\frac{\delta^r}{\delta\phi^*_i}G\right]u^A_i + \frac{1}{2}(-1)^{
\epsilon_i+1}\left[\frac{\delta^r}{\delta\phi^*_j}\frac{\delta^r}{\delta
\phi^*_i}G\right]\phi^*_kU^k_{ji} ~,\label{DeltanA}
\eeq
where the $U^k_{ij}$ are the structure coefficients for the supergroup
of transformations.\footnote{Taking for convenience that the supergroup
is semi-simple, with $(-1)^{\epsilon_i}U^i_{ij}=0$.}
They are related to the field transformations
$g^A(\phi',a)$ by the relation
\beq
\frac{\delta^r u^A_i}{\delta\phi^B}u^B_j - (-1)^{\epsilon_i\epsilon_j}
\frac{\delta^r u^A_j}{\delta\phi^B}u^B_i = -u^A_k U^k_{ij} ~,
\eeq
where
\beq
u^A_i(\phi) = \left.\frac{\delta^r g^A(\phi,a)}{\delta a^i}\right|_{a=0} ~.
\eeq

\noi
The $\Delta$-operator of eq. (\ref{DeltanA}) can be shown to be nilpotent
\cite{AD2}, and it gives rise to a new
non-Abelian antibracket by use of the relation (\ref{abdef}). Explicitly,
this antibracket takes the form \cite{AD2}
\beq
(F,G) \equiv (-1)^{\epsilon_i(\epsilon_A+1)}\frac{\delta^r F}{\delta\phi^*_i}
u^A_i\frac{\delta^l G}{\delta\phi^A} - \frac{\delta^r F}{\delta\phi^A}
u^A_i\frac{\delta^l G}{\delta\phi^*_i} + \frac{\delta^r F}{\delta\phi^*_i}
\phi^*_kU^k_{ij}\frac{\delta^l G}{\delta\phi^*_j} ~.
\eeq
In ref. \cite{AD2} this non-Abelian antibracket was derived directly in
the path integral (by integrating out the ghosts $c^A$), but it can readily
be checked that it is related to the associated $\Delta$-operator
(\ref{DeltanA}) in the manner expected from (\ref{abdef}). Because this
particular non-Abelian $\Delta$-operator is of 2nd order, the corresponding
antibracket automatically satisfies all the properties (3-6).

\noi
From the $\Delta$ operator we can derive the following master equation:
\beq
\Delta e^S=0\\
=\Delta S+\frac{1}{2}(S,S) \ \ , \epsilon(S)=0
\eeq
which is invariant under the following transformation:
\beq
\delta S=\Delta\epsilon+(\epsilon,S)
\eeq
where $\epsilon$ has ghost number equal to $-1$.

In the next section we will show a connection of this formalism to the
usual representation of gauge theories in term of forms.

\section{Connection to Gauge Theories}

It has already been noticed by Witten\cite{witten3} that the transformation
that leaves the master equation invariant is similar to the Yang-Mills 
transformations written in term of forms. In this context the master equation
itself can be thought as a sort of "field strength". In this section we will
show that this analogy can be very concrete indeed, providing us with an 
obvious
way to generalize Yang-Mills theories.

From now on what we called S in the previous sections will be denoted by $A$. 
This
is the " BV gauge field". It has $\epsilon(A)=0$ and will be, in general, a 
function
of the space coordinates $x^\mu$ and certain "internal variables" which can
be Grassmann even ($y^a$) or odd variables($y*_a$). Its exact nature will be 
described below.
The " BV gauge field" and some other fields that will be introduced soon have 
to be thought
as "superfields" and so can be expanded in a Taylor series in the internal 
variables.
 
Let us introduce the "field strength" F;
\beq
F=\Delta A+\frac{1}{2}(A,A), \label{13}
\eeq
Consider the transformation that leaves $F=0$ invariant:
\beqn
\delta A=\Delta\epsilon+(\epsilon,A)\\
\epsilon(\epsilon)=-1
\eeqn
These transformations form a closed algebra:
\beq
[\delta_\alpha,\delta_\beta]=\delta_{(\beta,\alpha)}
\eeq
It is easy to show that:
\beq
\delta F=(\epsilon, F), \label{17}
\eeq

Now, notice the crucial point that all these nice relations, equations
 (\ref{13}-\ref{17}), are based in only three properties : 

i)$\Delta^2=0$, 

ii)$\Delta$ is a second order
differential operator, both in the space time variables and in the internal 
variables and

iii)$\epsilon(A)=0$. mod 2.

i) and ii) implies that the antibracket derived from $\Delta$ using equation (2)
will satisfy the Jacobi identity equation (5).

To make contact with the Yang-Mills field we choose the following form of 
$\Delta$, which will acts on functions of $x^\mu$(space-time coordinates),
$\theta^\mu$ (an anticonmuting variable), $y^A$ and $y_a^*$(internal coordinates):
\beqn
\Delta=\theta^\mu\partial_\mu+\Delta_{NA}\\
\Delta_{NA} G \equiv (-1)^{\epsilon_i}\left[\frac{\delta^r}{\delta y^A}
\frac{\delta^r}{\delta y^*_i}G\right]u^A_i + \frac{1}{2}(-1)^{
\epsilon_i+1}\left[\frac{\delta^r}{\delta y^*_j}\frac{\delta^r}{\delta
y^*_i}G\right]y^*_kf^k_{ji} ~,\label{DeltaYM}\\
\epsilon(y^A)=0 \  \   \, \epsilon(y^*_a)=-1
\eeqn
$\partial_\mu$ means derivation with respect to $x^\mu$.$u_i^A(y)$ defines
the infinitesimal transformation of the internal bosonic variables $y^A$:
\beq
\delta y^A=u_i^a(y)\lambda^i
\eeq
$\lambda$ is an infinitesimal parameter. This transformation can be non-linear.

Here $\theta$($x^\mu$) anticonmute(conmute) with the Grassman odd(even)
variables in the internal sector. This garanties that $\Delta^2=0$. It can
be readely seen that $\Delta$ is a second order differential operator.
Notice that $\theta^\mu\partial_\mu$ must acts as a right derivative.

From this nilpotent operator, we derive the following antibracket:
\beq
(F,G) \equiv (-1)^{\epsilon_i(\epsilon_A+1)}\frac{\delta^r F}{\delta y^*_i}
u^A_i\frac{\delta^l G}{\delta y^A} - \frac{\delta^r F}{\delta y^A}
u^A_i\frac{\delta^l G}{\delta y^*_i} + \frac{\delta^r F}{\delta y^*_i}
y^*_kf^k_{ij}\frac{\delta^l G}{\delta y^*_j} ~.
\eeq

Now, let us make explicit the connection with the Yang-Mills field. Choose:
\beq
A=A^a_\mu(x)y^*_a\theta^\mu
\eeq
Since we must have $\epsilon(A)=0$, we 
asign $\epsilon(\theta)=1$ and $\epsilon(y^*_a)=-1$.
Then we get, from (\ref{13}-\ref{17}):
\beqn
\delta A^a_\mu(x)=\partial_\mu\epsilon^a+f^a_{cd}\epsilon^c A^d_\mu\\
F=(\partial_\mu A_\nu^a-\partial_\nu A_\mu^a-f^a_{cd}A^c_\mu A^d_\nu)
y^*_a\theta^\nu\theta^\mu\\
=F_{\mu\nu}^a y^*_a\theta^\nu\theta^\mu
\eeqn
which are the transformation and the field strength of the Yang-Mills
field.

We see that for a very particular $A$ satisfying the 
condition $\epsilon(A)=0$, we
already get the structure of the Yang-Mills field. But this 
condition can be
met by many more monomials in the $\theta,y^A,y^*_a$ variables, 
providing us
with generalizations of the Yang-Mills symmetry principle, 
through equations (\ref{13}-\ref{17}).

Before presenting some of these generalizations, let us prove a very 
useful
identity, corresponding to the Bianchi identity in the Yang-Mills case.

\section{Bianchi Identity}

Let us write $F$ in the following form:
\beq
F=e^{-A}\Delta e^A
\eeq
We get the following identity, using the nilpotency of $\Delta$:
\beq
\Delta(Fe^A)=0
\eeq
That is:
\beq
\Delta F +(F,A)=0
\eeq
This is the Bianchi identity.

\section{\sc Enhanced Gauge Symmetry}

\noi

Since the gauge transformations defined above form a closed algebra, 
they open the road to build gauge invariant Lagrangians involving 
interacting antisymmetric
tensors of arbitrary order. 

In this letter, we will consider in detail just the simplest 
generalization of Yang-Mills
provided by the Non-Abelian BV formalism. It will be clear how to 
proceed in
more complex situations.

Let us introduce the gauge field 
\footnote{The condition $\epsilon(A)=0$ permits also to incorporate 
additional 
terms, some of them 
containing odd monomials in $y^*_a$ and $\theta$, whose coefficients will be
fermionic fields. We will not discuss such fields here.}:
\beq
A=\phi+A^a_\mu\theta^\mu y^*_a+\frac{1}{2} A^{ab}_{\mu\nu}\theta_
\mu\theta_\nu y^*_a y^*_b
\eeq

We can compute the field strenght F:
\beqn
F=F_\mu\theta^\mu+\bar F_{\mu\nu}^a y^*_a\theta^\mu\theta^\nu+
\frac{1}{2}F_{\mu\nu\lambda}^{ab}
y^*_a y^*_b\theta^\mu\theta^\nu\theta^\lambda\\
F_\mu=\phi,_\mu\\
\bar F_{\mu\nu}^a=F_{\mu\nu}^a+ f^a_{ji} A^{ij}_{\mu\nu}\\
F_{\mu\nu\lambda}^{ab}=(A_{\nu\lambda,\mu}+ 
f^b_{ji} A^{ija}_{\mu\nu\lambda}+
 f^b_{ji} A^j_\mu A^{ai}_{\nu\lambda})_{ab;\mu\nu\lambda}
\eeqn
$()_{xy}$ means antisymmetrization with respects the indices $xy$.

Here $F_{\mu\nu}^a$ is the usual Yang-Mills field strength.

The gauge parameter is:
\beq
\epsilon=\epsilon^a(x)y^*_a+\frac{1}{2}\epsilon^{ab}_\mu y^*_a y^*_b
\theta^\mu
\eeq

We get the following gauge transformations of the component fields:
\beqn
\delta\phi=0\\
\delta A ^a_\mu (x)=\partial_\mu\epsilon^a(x) +
\epsilon^i(x) f^a_{ij} A^j_\mu(x) +\frac{1}{2}\epsilon^{ij}_\mu(x) f^a_{ji}\\
\delta A^{ab}_{\mu\nu}=\frac{1}{2}(\epsilon^{ab}_{\mu,\nu}-
\epsilon^{ab}_{\nu,\mu})-(f^a_{ji}\epsilon^j A^{ib}_{\mu\nu})_{ab}
+(f^a_{ji} A^i_\mu\epsilon^{jb}_\nu)_{ab,\mu\nu}
\eeqn

Also, we obtain:
\beqn
\delta F_\mu=0\\
\delta\bar F_{\mu\nu}^a=-f^a_{ji}\epsilon^j \bar F_{\mu\nu}^i\\
\delta F_{\mu\nu\lambda}^{kb}=(f^k_{jai}\epsilon^j F^{ib}_{\mu\nu\lambda}+
f^k_{ji}\epsilon^{jb}_\mu F_{\nu\lambda}^i)_{kb;\mu\nu\lambda}
\eeqn

It is easy to check that the following action is gauge invariant 
in four dimensions:
\beqn
S= \int d^4x \{a_1 \bar F_{\mu\nu}^i\bar F_{\mu\nu i} +a_2 
\tilde{\bar F_{\mu\nu}^i}\bar F_{\mu\nu i}+
a_3 A_{\mu\nu}^{ab} \epsilon^{\mu\nu\rho\lambda} 
\bar F^i_{\rho\sigma}f_{iab}\}
\eeqn

$\tilde{\bar F_{\mu\nu}^i}=
\epsilon_{\mu\nu\lambda\rho}\bar F_{\lambda\rho}^i$
is the dual of $\tilde{\bar F_{\mu\nu}^i}$.

To prove the invariance of the last term, we have to use the Bianchi 
identity.

$B_{\mu\nu}^i=A_{\mu\nu}^{ab} f_{ab}^i$ transforms exactly as the 
Freedman-Townsend two-form
does\cite{freedman} and when $\epsilon^a(x)=0$, 
the last term of our Lagrangian
is similar but not equal to the Freedman-Townsend action for the 
antisymmectric
tensor field. However this is not in contradiction with reference 
\cite{marc}
because even when $\epsilon^a(x)=0$, the Yang-Mills field transforms, 
according to equation(37).

\section{Characteristic Classes}

It is useful to introduce the following notation:
\beqn
F_2=\bar F_{\mu\nu}^at^a\theta^\mu\theta^\nu\\
F_3=F_{\mu\nu\lambda}^{ab}[t_a,t_b]\theta^\mu\theta^\nu\theta^\lambda
\eeqn

$t_a$ are the generators of the Lie algebra of the gauge group G,
satisfying:
\beq
[t_a,t_b]=if_{ab}^ct_c
\eeq

In term of these variables the transformation of the field strenghts are
 the following:
\beqn
\delta F_2=[\alpha,F_2]\\
\delta F_3=[\alpha, F_3]+[\alpha_1,F_2]\\
\alpha=\alpha^a t_a\\
\alpha_1=\alpha_\mu^{ab}[t_a,t_b]\theta^\mu
\eeqn
We readily check that the following object is gauge invariant:
\beq
C_k=\int d^{2k+3}x d^{2k+3}\theta tr F_2^kF_3
\eeq
Moreover, using the Bianchi identity:
\beqn
dF_2-\frac{1}{4} F_3+[F_2,A_1]=0\\
A_1=A_\mu^at_a\theta^\mu\\
d=\theta^\mu\partial_\mu
\eeqn
we get:
\beqn
d tr F_2^k=\frac{k}{4} tr F_2^{k-1}F_3
\eeqn
This implies that $C_k$ is a closed form. This is the analog of 
the Chern Class in
d=5,7,9.. dimensions.
On the other hand:
\beq
CS_k=\int d^{2k} x d^{2k} \theta tr F_2^k
\eeq
is also gauge invariant, giving the analog of the Chern-Simons class.

\section{Matter Fields}
\noi
To introduce matter fields in the formalism we borrow the transformation 
rule for
the field strength $F$:
\beq
\delta F=(\epsilon,F)
\eeq
It also forms a representation of the same closed algebra:
\beq
[\delta_\alpha,\delta_\beta]=\delta_{(\beta,\alpha)}
\eeq
This can be proven using the Jacobi identity.

From now on we will call "BV matter field" a function(superfield) 
$\phi$ of the space-time and
internal cocordinates whose transformation rule is:
\beq
\delta\phi=(\epsilon,\phi),\ \ \epsilon(\epsilon)=-1 \label{matter}
\eeq
$\epsilon(\phi)$ is arbitrary.

To built gauge invariant Lagrangians, involving BV matter fields, 
we need the concept of 
covariant derivative.

In the present formalism this is done using what is called in the 
BV formalism
the "quantum BRST transformation"
\beq
DG=\Delta G +(G,A)
\eeq

Indeed, under a gauge transformation $D\phi$ transforms as $\phi$ does:
\beqn
\delta A=\Delta\epsilon+(\epsilon,A)\\
\delta\phi=(\epsilon,\phi)
\eeqn
 implies
\beq
\delta(D\phi)=(\epsilon,D\phi)
\eeq
which is the fundamental property of the covariant derivative.

We also get:
\beq
D^2\phi=(\phi,F)
\eeq

So if $F$ vanishes $D$ is nilpotent.

We will say that $\Psi(x,y,\theta,y^*$ is fermionic
(bosonic) if  $\epsilon(\Psi(x,y,\theta,y^*))$ is 1(0), mod 2.

To make contact with the standard formulation of matter fields 
we consider the BV matter field,
\beq
\Psi(x,y,\theta,y^*)=\psi_A(x) y^A
\eeq
as an example.

Then we get using the equation (\ref{matter}),
\beq
\delta\psi_A=\epsilon^i\frac{\partial u_i^B}{\partial y^A}|_{y^A=0}\psi_B(x)
\eeq
which coincides with the gauge transformation of the matter field $\psi$ 
which belongs to 
the representation of the gauge group defined by $u$. For linear 
representations
of the gauge group $u_i^A$ is given by:
\beqn
\delta y^A=\lambda^i y^BT_{BA}^i\\
u_i^A=y^BT_{BA}^i\
\eeqn
$\lambda^i$ is an infinitesimal parameter and $T_{AB}^i$ are the Lie 
algebra generators.

Notice that the representation expanded by $u_i^A$ can be reducible, 
so this permits to incorporate
various fields transforming differently under the gauge group.

The Higgs field can be incorporated as in the last equation 
(starting with a scalar
BV matter field  and taking $u_i^A$ in the appropriate representation of G) 
or if it belongs into the
adjoint representation of the group, we can "unify" a fermion 
(in the $u_i^A$ representation
of G) and a scalar expanding the adjoint representation of G, 
choosing a fermion BV matter field:
 
\beqn
\Psi=\psi_A(x) y^A+\phi^a(x) y^*_a\\
\delta\psi_A=\frac{\partial u^a_B}{\partial y^A}|_{y^A=0}\epsilon^B\\
\delta\phi^a(x)=\epsilon^i f^a_{ij}\phi^j
\eeqn

The covariant derivative is, in terms of the components of the BV 
matter field:
\beqn
D_\mu\psi_A(x)=\partial_\mu\psi_A(x)+A^i_\mu T^i_{AB}\psi_B(x)\\
D_\mu\phi^k(x)=\partial_\mu\phi^k(x)-f_{ij}^k\phi^i(x)A^j_\mu
\eeqn

They coincide with the standard answer.

The simplest generalization of a matter field we can consider 
is the following:
\beqn
\Phi=\phi_A(x) y^A+\phi^a_{A\mu}(x)y^A y^*_a\theta^\mu \\
\delta\phi_A(x)=\epsilon^i \frac{\partial u^B_i}
{\partial y^A}|_{y^A=0}\phi_B\\
\delta\phi^a_{A\mu}=\epsilon^{ia}_\mu  \frac{\partial u^B_i}
{\partial y^A}|_{y^A=0}
\phi_B+\epsilon^i f^a_{ij}\phi^j_{A\mu}
\eeqn

The covariant derivative is, in this case:
\beqn
D_\mu\phi=\partial_\mu\phi+\phi_{A\mu} u^A_i-\phi_A u^A_i A^i_\mu\\
\phi=\phi_A y^A\\
\phi^i_\mu=\phi^i_{A\mu} y^A
\eeqn

From the transformation law of the matter field, we get that
\beq
D_\mu\phi_A D_\mu\phi_B g^{AB}
\eeq
is gauge invariant.$g^{AB}$ ia an invariant tensor under 
transformations of G.

\section{Conclusions}

In this letter, we have presented a particular realization of the 
Non-Abelian BV
formalism, which incorporates in a unified way all the ingredients of 
the standard
model:Yang-Mills field, covariant derivatives, Higgs fields and fermions. 

The whole structure offer ample scope for generalizations: By considering
superfields in the internal($y^A,y^*_a$ and external variables $\theta^\mu$
with suitable Grassman signatures, we can incorporate, gauge bosons of
higher spin interacting with fermions and bosons expanding arbitrary 
representations
of the Lorentz group. This is done in a systematic form, by 
enlarging the symmetry
of the models.

We have offered some simple examples of these generalizations. 
In particular
we have shown how to built Characteristic Classes involving a 
non-abelian
two-form. We also constructed a gauge invariant Lagrangian 
coupling the two-form gauge field
to the Yang-Mills field in four space-time dimensions. This 
differs from the Freedman-Townsend coupling, although the 
transformation law for the two-form gauge field is the same.
The reason being that the one-form gauge field transforms
differently.

\section*{Acknowledgements}

The author wants to thank M. Henneaux for a conversation.
His work has been partially supported by Fondecyt \# 1950809, 
a Conacyt(M\'exico)-Conicyt(Chile) collaboration and a CNRS-Conicyt
joint project.


\begin{thebibliography}{88}

\bibitem{witten1} For a modern review of String Theory see:
     M.B. Green,J.H. Schwarz and E. Witten, "Superstring Theory"
     vol 1,2 , Cambridge University Press 1987.
\bibitem{witten2} E. Witten, Nucl. Phys. B268(1986)253.
\bibitem{zwiebach}E. Witten and B. Zwiebach, Nucl. Phys. {\bf B377}
(1992) 55.\newline B. Zwiebach, Nucl. Phys. {\bf B390} (1993) 33.
\bibitem{BV}I.A. Batalin and G.A. Vilkovisky, Phys. Lett.
{\bf 102B} (1981) 27; Phys. Rev. {\bf D28} (1983) 2567 [E: {\bf D30}
(1984) 508]; Nucl. Phys. {\bf B234} (1984) 106; J. Math. Phys.
{\bf 26} (1985) 172.\newline 
For some reviews see: M. Henneaux, Nucl. Phys. B (Proc. Suppl.) {\bf 18A}
(1990) 47.
\newline J.M.L. Fisch, Univ. Brux. preprint ULB TH2/90-01. \newline
M. Henneaux and C. Teitelboim, "Quantization of Gauge Systems",
Princeton University Press, Princeton, New Jersey (1992).


\bibitem{AD}J. Alfaro and P.H. Damgaard, Nucl. Phys. {\bf B404}
(1993) 751.
\bibitem{2} J. Alfaro and P.H. Damgaard, Phys. Lett. B334(1994)369.

\bibitem{AD2} J. Alfaro and P.H. Damgaard, 
Nuclear Physics B455(1995)409.

\bibitem{AD1} J. Alfaro and P.H. Damgaard,
Phys. Lett. {\bf B369} (1996) 289.

\bibitem{last} K. Bering, P. H. Damgaard, J. Alfaro,
Nucl.Phys.B478(1996)459. 
\bibitem{sen} A. Sen, "Some applications of String Field theory", 
TIFR/TH/91-39,
September 1991.

\bibitem{freedman} D. Freedman and P.K. Townsend, Nucl. Phys. 
B177(1981)282.

\bibitem{marc} M. Henneaux,  Phys.Lett.B368(1996)83.

\bibitem{AD3}J. Alfaro and P.H. Damgaard, Phys. Lett. {\bf B222}
(1989) 425; Ann. Phys. (NY) {\bf 202} (1990) 398.\newline
J. Alfaro, P.H. Damgaard, J. Latorre and
D. Montano, Phys. Lett. {\bf B233} (1989) 153.

\bibitem{witten3}E. Witten, Mod. Phys. Lett. {\bf A5} (1990) 487.

\end{thebibliography}
\end{document}